\documentclass[%
 reprint,
 superscriptaddress,
 amsmath,amssymb,
 aps,
]{revtex4-2}

\usepackage{graphicx}
\usepackage{dcolumn}
\usepackage{bm}
\usepackage[dvipsnames]{xcolor}

\begin{document}

\preprint{APS/123-QED}

\title{Unveiling the Spin--Valley Structure of Dipolar Exciton Ladders in R-stacked WSe$_2$/WS$_2$ Moir\'e Heterobilayers}

\author{Byeong Wook Cho}
\thanks{These authors contributed equally to this work.}
\affiliation{%
 Institute of Photonics and Quantum Sciences, SUPA, Heriot-Watt University, Edinburgh, UK.
}
\author{Tatyana V. Ivanova}
\thanks{These authors contributed equally to this work.}
\affiliation{%
 Institute of Photonics and Quantum Sciences, SUPA, Heriot-Watt University, Edinburgh, UK.
}

\author{Zhe Li}
\affiliation{%
 Institute of Photonics and Quantum Sciences, SUPA, Heriot-Watt University, Edinburgh, UK.
}

\author{Takashi Taniguchi}
\affiliation{\TsukubaTakashi}

\author{Kenji Watanabe}
\affiliation{\TsukubaKenji}

\author{Brian D. Gerardot}
\email{B.D.Gerardot@hw.ac.uk}
\affiliation{%
 Institute of Photonics and Quantum Sciences, SUPA, Heriot-Watt University, Edinburgh, UK.
}

\author{Mauro Brotons-Gisbert}
\email{m.brotons\_i\_gisbert@hw.ac.uk}
\affiliation{%
 Institute of Photonics and Quantum Sciences, SUPA, Heriot-Watt University, Edinburgh, UK.
}

\newcommand{\TsukubaKenji}{Research Center for Electronic and Optical Materials, National Institute for Materials Science, 1-1 Namiki, Tsukuba 305-0044, Japan}
\newcommand{\TsukubaTakashi}{Research Center for Materials Nanoarchitectonics, National Institute for Materials Science,  1-1 Namiki, Tsukuba 305-0044, Japan}
\date{\today}

\begin{abstract}
Localized interlayer excitons in moir\'e heterobilayers can form dipolar exciton ladders, yet their internal spin--valley structure remains unresolved. 
Here, we use helicity-resolved magneto-photoluminescence to identify the microscopic origin of the ladder in R-stacked WSe$_2$/WS$_2$ at charge neutrality and one-electron filling of the moir\'e lattice. 
At charge neutrality, the first two emission peaks correspond to a spin-triplet interlayer exciton and a triplet--triplet two-exciton state separated by 38 meV, reflecting the on-site dipolar interaction.
The opposite Zeeman response of the apparent third rung of the ladder rules out its assignment as a spin-conserving three-exciton state and instead identifies it as a triplet--singlet two-exciton configuration with a 22 meV offset set by the WS$_2$ conduction-band spin splitting. 
At one-electron filling, the correlated electronic background gives rise to charged one- and two-exciton states and intervalley/intravalley two-exciton configurations, while reducing the effective exciton--exciton interaction.
Our results establish a spin--valley-resolved picture of dipolar exciton ladders beyond simple occupation-number physics in moir\'e heterobilayers.
\end{abstract}

\maketitle
Moir\'e heterobilayers of transition-metal dichalcogenides (TMDs) provide a tunable platform for localizing interlayer excitons in periodic nanoscale potentials~\cite{tang2020simulation-64e,wu2018hubbard-675,jin2019observation-fdb,regan2020mott-1b0,liu2021excitonic-783}.
In type-II heterobilayers such as MoSe$_2$/WSe$_2$ and WSe$_2$/WS$_2$, electrons and holes reside in different layers, forming long-lived interlayer excitons (IXs) with permanent out-of-plane electric dipoles~\cite{rivera2015observation-8c4,xiong2023correlated-a7f,brotons-gisbert2021moir-trapped-e2f,liu2021signatures-1b1,yuan2020twist-angle-dependent-439,zhang2022correlated-264,fang2023localization-0dd,wang2021moir-87c,frg2019cavity-control-fee,qian2024lasing-eba,liu2019room-cf9,zhao2023excitons-946,tagarelli2023electrical-c77,wilson2017determination-cd6,troue2023extended-4d9,jauregui2019electrical-d0c}.
Moir\'e confinement makes these dipolar excitons optically addressable in both isolated and interacting regimes: at low exciton density, individual moir\'e traps can host isolated IXs that behave as quantum-dot-like emitters, whereas at higher density, multiple IXs can occupy and interact within the same moir\'e potential~\cite{seyler2019signatures-e9b,baek2020highly-ba8,yu2017moir-cb3,brotons-gisbert2021moir-trapped-e2f,liu2021signatures-1b1}.
This combination of optical access, spatial confinement, and strong dipolar interactions makes moir\'e-trapped IXs a natural platform for exploring few- and many-body excitonic states.

In this high-density regime, dipolar interactions become prominent in the photoluminescence (PL) spectrum. 
The long-range component of the dipolar interaction continuously blueshifts the IX resonances, whereas the on-site component gives rise to discrete higher-energy emission peaks associated with successive IX occupation of a single moir\'e site, commonly interpreted as a dipolar exciton ladder~\cite{park2023dipole-b29,lian2024valley-polarized-86a,li2020dipolar-cc0,huang2025moir-orbital-resolved-748}.
In WSe$_2$/WS$_2$ moir\'e heterobilayers, the on-site Hubbard interaction between moir\'e-confined IXs reaches $\sim$30--40~meV, far exceeding that reported in MoSe$_2$/WSe$_2$ systems ($\sim$2~meV)~\cite{park2023dipole-b29,li2020dipolar-cc0}. 
This exceptionally large interaction energy makes WSe$_2$/WS$_2$ well suited for investigating excitonic Bose--Hubbard physics~\cite{yu2017moir-cb3,gtting2022moir-bose-hubbard-eb4,gao2024excitonic-2e9}.
Beyond its large interaction scale, the observed dipolar ladder also depends strongly on stacking geometry. 
Previous optical studies reported up to three resolved IX peaks in R-stacked samples, whereas only two were resolved in H-stacked configurations~\cite{park2023dipole-b29,lian2024valley-polarized-86a}. 
This contrast has been attributed to their different real-space character: R-stacked IXs are vertically aligned and localized at a single registry, whereas H-stacked IXs are laterally displaced over multiple registries, leading to a more delocalized IX wavefunction~\cite{lian2024valley-polarized-86a,devenica2026collective-1d2}.
However, the unequal energy spacing of the three peaks observed in R-stacked WSe$_2$/WS$_2$ cannot be explained by on-site dipolar repulsion alone, indicating that additional internal degrees of freedom must be involved~\cite{park2023dipole-b29,lian2024valley-polarized-86a}.

Moreover, moir\'e-trapped IXs interact with correlated carriers introduced by electrostatic doping~\cite{miao2021strong-d88,kim2024correlation-driven-634,wang2023intercell-416,chen2022tuning-e35}.
The resulting carrier--exciton interactions can shift IX resonance energies and generate additional excitonic complexes, whose microscopic nature remains poorly understood.
A key missing ingredient in understanding both the charge-neutral and doped regimes is the internal spin--valley configuration of these IXs, inherited from the constituent TMD monolayers~\cite{xiao2012coupled-aab,mak2012control-e01,xu2014spin-bf1}.
Because the accessible spin-split transitions and valley-dependent optical selection rules are determined by both the spin--valley configurations of the constituent TMD monolayers and the local atomic registry of the moir\'e site, multiple spin--valley-distinct IX states can contribute to the observed emission spectrum~\cite{brotons-gisbert2021moir-trapped-e2f,yu2018brightened-a58,wang2020giant-c39}.
Resolving the spin--valley configurations of these excitonic states in R-stacked WSe$_2$/WS$_2$ is therefore essential to understand both the unequal spacing of the dipolar ladder at charge neutrality and the emergence of additional resonances in the presence of correlated carriers.

In this Letter, we present a helicity-resolved magneto-optical study of moir\'e-confined IXs in R-stacked WSe$_2$/WS$_2$ heterobilayers as a function of exciton density and carrier filling $\nu$.
Here, $\nu$ denotes the number of excess carriers per moir\'e unit cell, with positive and negative values corresponding to electron and hole doping, respectively.
At charge neutrality, we show that the exciton ladder cannot be understood as a simple sequence of spinless dipolar states.
Instead, it comprises a spin-triplet ground-state interlayer exciton and two higher-energy two-exciton states with triplet--triplet and triplet--singlet spin--valley configurations. 
This microscopic assignment naturally explains the unequal ladder spacing through the spin-split WS$_2$ conduction band.
At $\nu=1$, the correlated electronic background reshapes the PL spectrum, giving rise to charged exciton--carrier complexes, charged two-exciton states, and a fine structure of intervalley and intravalley two-exciton configurations.
Moreover, we observe a reduction of the effective inter-exciton dipolar repulsion at $\nu=1$, demonstrating that resident electrons renormalize the interaction strength of bosonic excitons in the moiré lattice.

We fabricated three R-stacked WSe$_2$/WS$_2$ heterobilayer devices, R1--R3, using a dry-transfer method.
The relative twist angles between WSe$_2$ and WS$_2$ were determined by polarization-resolved second-harmonic generation (SHG) measurements (see Supplemental Material, Sec.~S1 and Fig.~S1).
All data presented in the main text were acquired from device R1.
The devices employ a dual-gated geometry, enabling independent control of charge doping and out-of-plane electric field (see Supplemental Material, Sec.~S2).
PL measurements were performed at 3.9~K using a continuous-wave excitation laser at 1.680~eV, resonant with the 1$s$ bright exciton state of WSe$_2$, unless otherwise specified (see Supplemental Material, Fig.~S2).
Figure~1(a) shows a schematic of a dual-gated R-stacked WSe$_2$/WS$_2$ heterobilayer where electrons and holes are vertically aligned and localized at the same R$_{\mathrm{h}}^{X}$ moir\'e site, forming moir\'e-confined IXs with permanent out-of-plane dipole moments~\cite{wang2023intercell-416,park2023dipole-b29}. 
Here, R$_{\mathrm{h}}^{X}$ denotes the R-type stacking with the chalcogen ($X$) site of the electron layer vertically aligned with the hexagon center ($h$) of the hole layer. 
This contrasts with H-stacked WSe$_2$/WS$_2$, where the ground-state IX has a more extended real-space structure: one electron (hole) is localized near a moir\'e trap, while the wave function of the opposite hole (electron) is distributed over three symmetry-equivalent neighboring traps, giving the exciton an in-plane quadrupolar character in addition to its vertical dipole~\cite{wang2023intercell-416,devenica2026collective-1d2}.
In both stacking configurations, a single moir\'e site can host multiple IXs, which appear in PL spectra as successive emission peaks separated by the on-site dipolar repulsion, $U_{dd}$, with increasing exciton density.
It has been suggested that the R-stacked WSe$_2$/WS$_2$ can support up to three IXs in a single moir\'e unit cell whereas the H-stacked configuration supports only two owing to its relatively delocalized IX nature~\cite{park2023dipole-b29,lian2024valley-polarized-86a}. 
Figure~1(b) illustrates the relevant IX transitions in the R-stacked configuration, where the $+K$ ($-K$) valleys of WSe$_2$ and WS$_2$ are nearly aligned.
The ground-state transition corresponds to the spin-triplet interlayer exciton (IX$^{(\mathrm{T})}$), while transitions involving the upper spin-split conduction band give rise to spin-singlet interlayer exciton (IX$^{(\mathrm{S})}$). 
These two states are separated by the conduction-band spin splitting, $\Delta_{\mathrm{C}}$ (around 16--30~meV), and exhibit opposite circular polarization selection rules~\cite{eickholt2018spin-991,absor2016strain-controlled-ea0,komider2013large-cca,yu2018brightened-a58} (see Supplemental Material, Sec.~S3).

\begin{figure}[!t]
    \centering
    \includegraphics[width=0.5\textwidth]{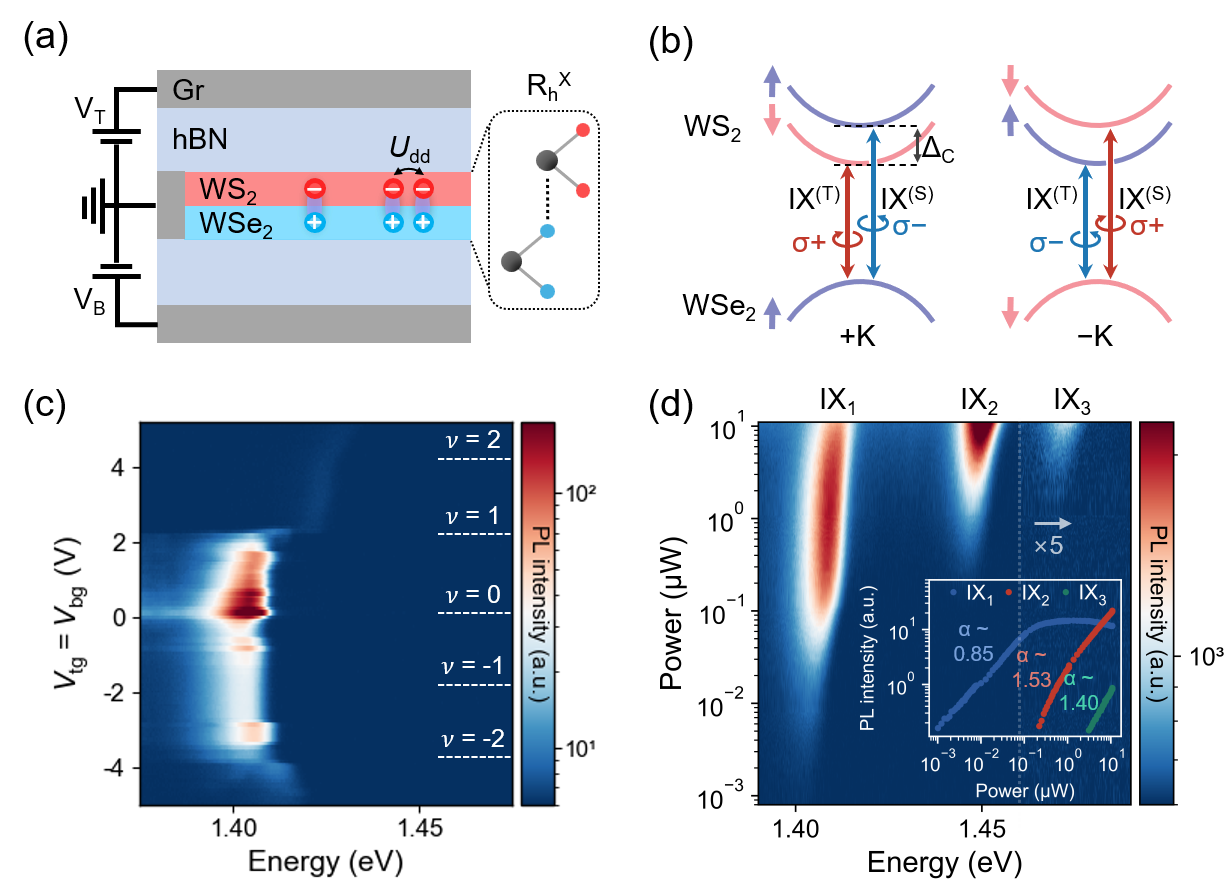}
    \caption{
    (a) Schematic of the hBN-encapsulated dual-gated heterobilayer, where moir\'e-confined IXs carry out-of-plane electric dipoles and interact through on-site dipolar repulsion $U_{\mathrm{dd}}$.
    (b) Spin--valley configurations and helicity-dependent optical selection rules for triplet IX$^{(\mathrm{T})}$ and singlet IX$^{(\mathrm{S})}$ transitions, separated by the WS$_2$ conduction-band spin splitting $\Delta_{\mathrm{C}}$.
    (c) Filling-dependent PL spectra measured at 10~nW under linearly polarized 1.680~eV excitation, showing systematic modulation of the IX emission with moir\'e lattice filling.
    (d) Excitation-power-dependent PL spectra at charge neutrality under the same excitation energy, showing the emergence of IX$_2$ and IX$_3$ above IX$_1$.
    The high-energy spectral region is multiplied by a factor of five for clarity.
    Inset: integrated PL intensities of IX$_1$, IX$_2$, and IX$_3$ with power-law fits.
    }
    \label{fig:fig1}
\end{figure}
\begin{figure*}[!t]
    \centering
    \includegraphics[width=0.85\textwidth]{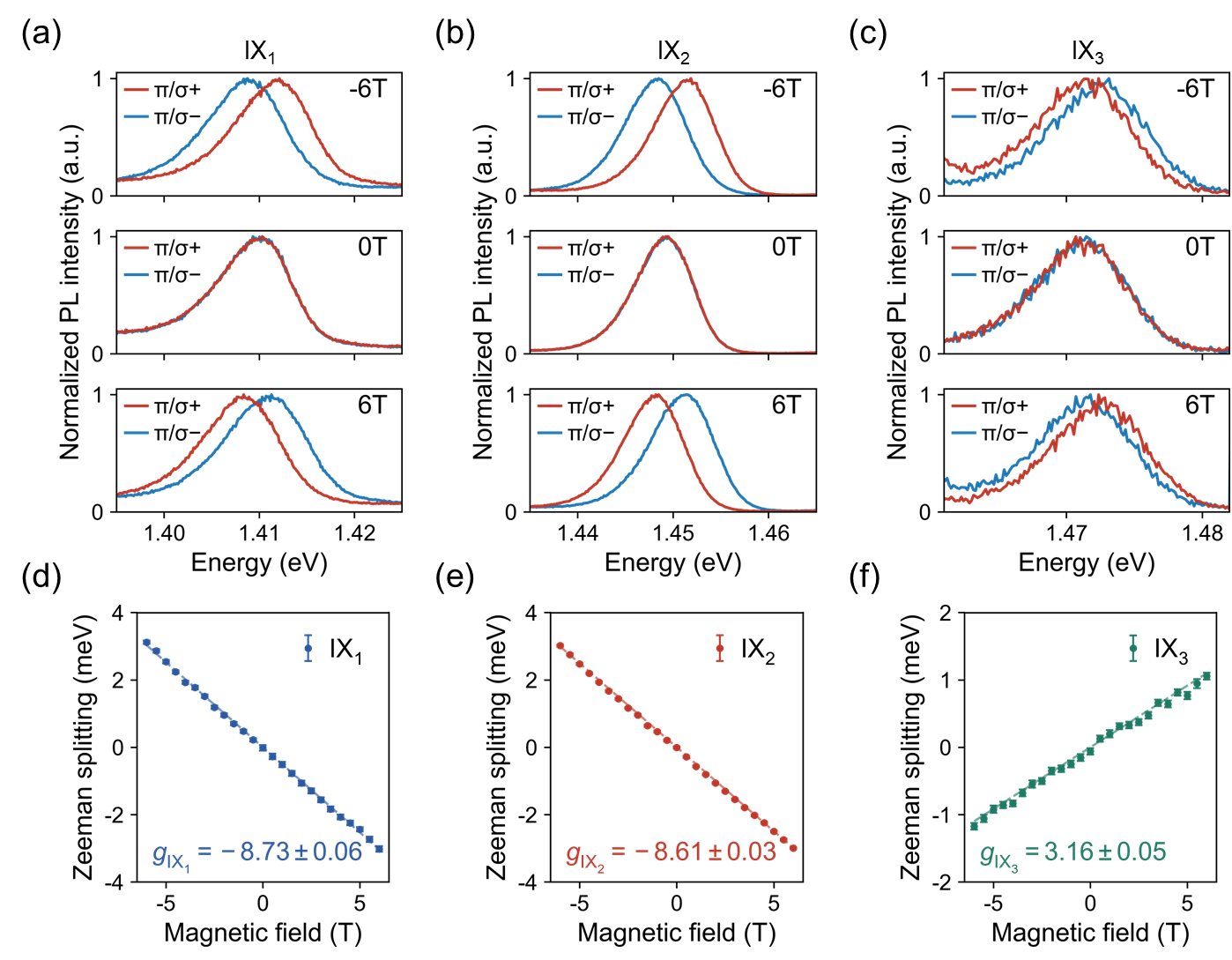}
    \caption{
    (a--c) Helicity-resolved PL spectra of IX$_1$, IX$_2$, and IX$_3$ at representative magnetic fields ($B=-6$, 0, and $+6$~T), under linearly polarized 1.72~eV excitation with a power of 20~$\mu$W.
    (d--f) Extracted Zeeman splitting, defined as $\Delta E = E_{\sigma^+}-E_{\sigma^-}$, as a function of magnetic field for IX$_1$, IX$_2$, and IX$_3$.
    Dashed lines are linear fits using $\Delta E = g\mu_{\mathrm{B}}B$.
    }
    \label{fig:fig2}
\end{figure*}

Figure~1(c) shows the filling-dependent PL spectra measured at an excitation power of 10~nW.
The PL spectrum is dominated by the single-IX peak IX$_1$, whose intensity and peak energy systematically change with moir\'e lattice filling.
These lattice-filling-dependent PL modulations demonstrate the gate tunability of device R1.
Here, we focus first on the charge-neutral regime, while carrier-induced modifications of the IX spectrum are discussed later in Fig.~4.
Figure~1(d) shows the excitation power-dependent PL spectrum at charge neutrality ($\nu=0$).
At low excitation power, the emission is dominated by a single peak (IX$_1$) at 1.40 eV with a narrow linewidth of 11~meV (see Supplemental Material, Fig.~S3).  
With increasing excitation power, IX$_1$ exhibits a continuous blueshift and its linewidth becomes narrower down to 8~meV at 140~nW.
The blueshift is consistent with increasing long-range dipolar repulsion between IXs, while the linewidth narrowing suggests reduced disorder broadening as the exciton population approaches a bosonic Mott-like ordered configuration with nearly one IX per moir\'e site ~\cite{lian2024valley-polarized-86a}. 
Above 140~nW, the linewidth of IX$_1$ broadens and a higher-energy peak, IX$_2$, emerges 38~meV above IX$_1$ (see Supplemental Material, Fig.~S3).
We assign IX$_2$ to a two-IX state localized at the same moir\'e site.
This assignment is supported by the excitation-power dependence of the integrated PL intensities~\cite{wang2019optical-0ef}, shown in the inset of Fig.~1(d): IX$_1$ exhibits a sublinear scaling with an exponent of 0.85 and saturates near the onset of IX$_2$, whereas IX$_2$ shows a superlinear scaling with an exponent of 1.53.
The energy separation between IX$_2$ and IX$_1$, denoted $\Delta E_{21}=38$~meV, therefore represents the energy cost for two IXs to occupy the same moir\'e site, corresponding to the on-site dipolar repulsion $U_{dd}$.
At higher excitation powers ($3~\mu$W), a third emission peak (IX$_3$) appears near 1.47~eV.
Although IX$_3$ has previously been attributed to a three-exciton state within a single moir\'e site, its separation from IX$_2$, $\Delta E_{32}=22$~meV, is substantially smaller than $\Delta E_{21}=38$~meV\cite{park2023dipole-b29,lian2024valley-polarized-86a}.
This reduced spacing is inconsistent with a simple dipolar ladder in which IX$_3$ represents a three-IX state.
Moreover, the power-law exponent of IX$_3$ is 1.40, slightly smaller than that of IX$_2$.
We therefore attribute IX$_3$ to a two-exciton state with a spin--valley configuration distinct from that of IX$_2$.
Similar power-dependent intensity evolution and energy hierarchy of IX$_1$--IX$_3$ are observed in devices R2 and R3 (see Supplemental Material, Fig.~S4).

\begin{figure*}[!t]
    \centering
    \includegraphics[width=0.95\textwidth]{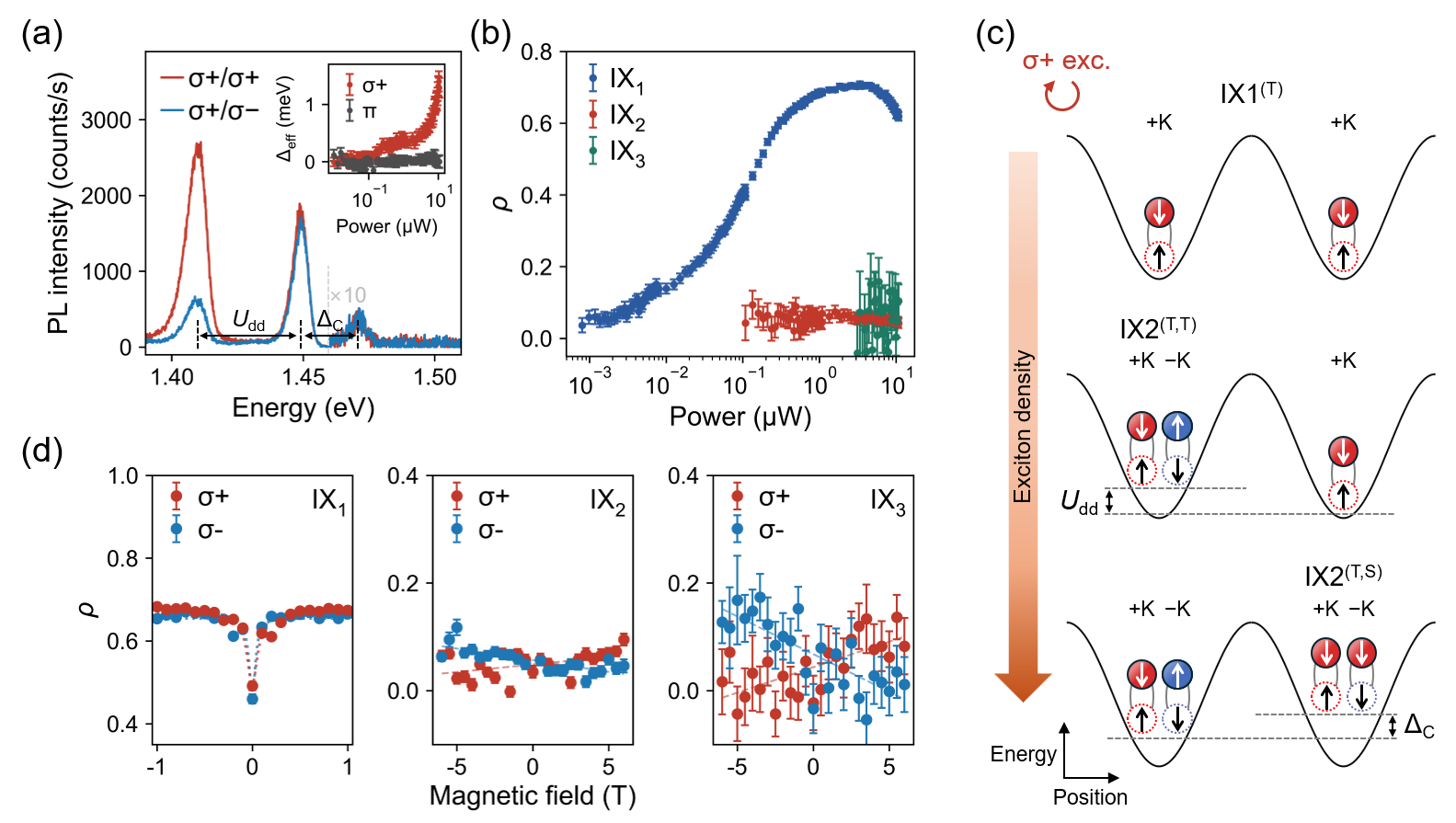}
    \caption{
    (a) Helicity-resolved PL spectra measured under $\sigma^+$ excitation with a power of 7~$\mu$W.
    The high-energy IX$_3$ region is multiplied by a factor of 10 for clarity.
    The inset compares the IX$_1$ splitting under linearly polarized and $\sigma^+$ excitation, highlighting the excitation-helicity-induced energy splitting.
    (b) Degree of circular polarization ($\rho$) of IX$_1$, IX$_2$, and IX$_3$ as a function of excitation power.
    (c) Schematic potential-energy landscape of the IX population dynamics under $\sigma^+$ excitation at charge neutrality.
    Increasing excitation density enhances the valley imbalance of IX$_1^{(\mathrm{T})}$ and is accompanied by the formation of intervalley two-exciton states, yielding IX$_2^{(\mathrm{T,T})}$ and IX$_2^{(\mathrm{T,S})}$ with negligible net valley polarization.
    (d) Magnetic-field dependence of $\rho$ for IX$_1$, IX$_2$, and IX$_3$ under $\sigma^+$ and $\sigma^-$ excitation.
    }
    \label{fig:fig3}
\end{figure*}

We performed helicity-resolved PL measurements under out-of-plane magnetic fields to investigate the origin of IX$_3$. 
Figures~2(a)--2(c) show normalized $\sigma^+$- and $\sigma^-$-resolved PL spectra of IX$_1$, IX$_2$, and IX$_3$, respectively, measured under linearly polarized 1.72~eV excitation at representative out-of-plane magnetic fields ($B=-6$, 0, and $+6$~T; see Supplemental Material, Fig.~S5 for the full magnetic-field-dependent spectra).
The applied magnetic field lifts the valley degeneracy through the Zeeman effect, producing opposite energy shifts for excitonic transitions in the $+K$ and $-K$ valleys and thereby enabling identification of their spin--valley configurations~\cite{ciarrocchi2019polarization-195,macneill2014breaking-96c}.
For each magnetic field, we extract the fitted PL peak energies $E_{\sigma^+}$ and $E_{\sigma^-}$ from the $\sigma^+$ and $\sigma^-$ detection channels, respectively.
The Zeeman splitting is then defined as $\Delta E = E_{\sigma^+} - E_{\sigma^-} = g\mu_B B$, where $g$ is the effective exciton $g$-factor and $\mu_B$ is the Bohr magneton.
Figures~2(d)--2(f) show the extracted energy splittings versus magnetic field, together with linear fits. 
IX$_1$ exhibits a $g$-factor of $g_{\mathrm{IX}_1} = -8.73 \pm 0.06$, consistent with previously reported values for the IX$^{(\mathrm{T})}$ in R-stacked WSe$_2$/WS$_2$~\cite{wu2025highly-80a}. 
IX$_2$ shows a nearly identical $g$-factor of $g_{\mathrm{IX}_2} = -8.61 \pm 0.03$, indicating that the corresponding two-IX state retains the same triplet optical transition.
In contrast, IX$_3$ exhibits a $g$-factor of $g_{\mathrm{IX}_3} = 3.16 \pm 0.05$, with opposite sign and significantly reduced magnitude. 
This value is consistent with expectations for the IX$^{(\mathrm{S})}$, as estimated within a single-particle framework including spin, atomic orbital, and Berry-curvature contributions (see Supplemental Material, Sec.~S4 and Fig.~S6)~\cite{brotons-gisbert2020spinlayer-e16,xiao2012coupled-aab}.
The same single-particle framework also captures the $g$-factors of IX$_1$ and IX$_2$, supporting the assignment of IX$_3$ to a transition involving IX$^{(\mathrm{S})}$.
Together with the nearly quadratic power dependence relative to IX$_1$ and the energy separation $\Delta E_{32}$ comparable to the WS$_2$ conduction-band spin splitting $\Delta_{\mathrm{C}}$, these results indicate that IX$_3$ originates from a two-exciton state composed of one IX$^{(\mathrm{T})}$ and one IX$^{(\mathrm{S})}$ confined in the same moir\'e site.

Figure~3(a) compares PL spectra measured at $7~\mu$W in co- and cross-circular excitation--detection configurations, providing access to the spin--valley configurations of the observed IX peaks. 
The polarization response of the optical setup was calibrated to remove systematic helicity-dependent detection asymmetries (see Supplemental Material, Sec.~S5 and Fig.~S7). 
As reported previously~\cite{wu2025highly-80a,lian2024valley-polarized-86a,park2023dipole-b29}, IX$_1$ exhibits strong co-circular polarization with respect to the excitation helicity (Fig.~3(b)).
We quantify the degree of circular polarization as $\rho=(I_{\mathrm{co}}-I_{\mathrm{cross}})/(I_{\mathrm{co}}+I_{\mathrm{cross}})$, where $I_{\mathrm{co}}$ and $I_{\mathrm{cross}}$ are the integrated PL intensities of the corresponding IX peak in the co- and cross-circular detection channels, respectively.
For IX$_1$, $\rho$ increases with excitation power and approaches 0.7 near 2~$\mu$W.
We also observe a small energy splitting, $\Delta_{\mathrm{eff}}=E_{\mathrm{co}}-E_{\mathrm{cross}}$, between the co- and cross-circular emission channels of IX$_1$, with $E_{\mathrm{co}}>E_{\mathrm{cross}}$.
This splitting increases with excitation power and is attributed to a valley population imbalance between the $+K$ and $-K$ valleys~\cite{li2021optical-e48} (see the inset of Fig.~3(a)).
This helicity-induced energy mismatch between the two valley configurations can suppress exchange-mediated valley mixing, leading to an enhanced $\rho$ of IX$_1$ at elevated excitation powers.
At excitation powers above 5~$\mu$W, the IX$_1$ polarization gradually decreases, which we attribute to exciton dissociation and enhanced intervalley scattering at high IX densities. 
This reduction is consistent with the decrease of the IX$_1$ PL intensity in the same high-power regime (see the inset of Fig.~1(d)).
In contrast, $\rho$ for IX$_2$ remains nearly zero over the investigated excitation-power range.
This nearly vanishing circular polarization is consistent with our assignment of IX$_2$ to an intervalley two-exciton state composed of one IX$^{(\mathrm{T})}$ in the $+K$ valley and the other IX$^{(\mathrm{T})}$ in the $-K$ valley, or vice versa, for which the two circular emission pathways are nearly indistinguishable~\cite{park2023dipole-b29}.
Such an intervalley configuration is energetically favored because the exchange interaction is reduced for anti-aligned spin--valley configurations. 
We further note that $\rho$ for IX$_1$ increases more rapidly above the onset of IX$_2$.
This observation is consistent with a scenario in which IX$_2$ forms preferentially from opposite-valley excitons, thereby reducing the cross-polarized IX population and increasing $\rho$ for IX$_1$~\cite{wu2025highly-80a}.
Similarly, IX$_3$ exhibits a nearly zero $\rho$ over the investigated excitation-power range, indicating that it also involves an intervalley two-IX configuration rather than a co-circularly polarized intravalley state.
Together with the opposite-sign $g$-factor of IX$_3$ discussed above, the nearly vanishing $\rho$ supports its assignment to an intervalley triplet--singlet two-IX configuration.
Based on these observations, we identify IX$_1$, IX$_2$, and IX$_3$ as a triplet interlayer exciton, IX$1^{(\mathrm{T})}$, a triplet--triplet two-exciton state, IX$2^{(\mathrm{T,T})}$, and a triplet--singlet two-exciton state, IX$2^{(\mathrm{T,S})}$, respectively (Fig.~3(c)). 
Here, the number following IX denotes the number of constituent excitons, while the superscripts denote their spin character.

Figure~3(d) shows the $\rho$ of IX$_1$, IX$_2$, and IX$_3$ as a function of out-of-plane magnetic field. 
The polarization of IX$_1$ was measured at 150~nW, for which the zero-field polarization remains unsaturated.
For both $\sigma^+$ and $\sigma^-$ excitation, the polarization increases sharply with magnetic field, reaching $0.7$ at $B_c \approx 0.05$ T before saturating.
This characteristic field is comparable to values previously reported for R-stacked WSe$_2$/WS$_2$ heterobilayers~\cite{wu2025highly-80a} (see Supplemental Material, Sec.~S6 for the fitting procedure).
This behavior is attributed to the magnetic-field-induced lifting of the $+K$/$-K$ valley degeneracy, which suppresses electron--hole-exchange-driven depolarization~\cite{yu2014valley-b1a,wu2025highly-80a}.
A residual depolarization of about 0.3 persists even at high magnetic fields and high excitation powers, while the saturation field remains similar across the investigated excitation-power range (see Supplemental Material, Fig.~S8).
This residual depolarization is likely due to the long IX lifetime, which allows depolarizing channels to remain effective even when exchange-induced valley mixing is partially suppressed~\cite{jin2018ultrafast-bb4,jiang2021interlayer-bce}.
Since an out-of-plane magnetic field lifts the valley degeneracy, it could in principle modify the valley population balance and thus the polarization of intervalley two-exciton states. 
However, IX$_2$ shows negligible $\rho$ throughout the investigated magnetic-field range, indicating that its two intervalley emission pathways remain nearly symmetric.

By contrast, IX$_3$ exhibits a field- and excitation-helicity-dependent polarization response.
In our assignment, IX$_3$ arises from recombination of the spin-singlet component of an intervalley two exciton configuration, composed of IX$^{(\mathrm{T})}$ in one valley and IX$^{(\mathrm{S})}$ in the opposite valley (Fig.~3(c)).
According to the selection rules in Fig.~1(b), $\sigma^+$ ($\sigma^-$) excitation addresses the IX$^{(\mathrm{S})}$ transition in the $-K$ ($+K$) valley.
The observed helicity dependence therefore indicates that the higher-energy valley branch develops a larger $\rho$ than the lower-energy branch.
A related behavior has been reported for bright spin-singlet excitons in monolayer WSe$_2$, where an out-of-plane magnetic field modifies the valley-dependent exciton dispersion through electron--hole exchange interactions~\cite{aivazian2015magnetic-9f2}.
In that case, the magnetic field lifts the valley degeneracy and separates the exchange-coupled exciton branches, favoring radiative recombination from the upper branch.
Although the spatially indirect character of IX$_3$ is expected to substantially reduce both the exchange interaction and oscillator strength, the spin-singlet recombination channel may retain a similar exchange-mediated polarization mechanism.
Such a scenario could account for the distinct magnetic-field-dependent polarization response of IX$_3$.

\begin{figure*}[!t]
    \centering
    \includegraphics[width=0.95\textwidth]{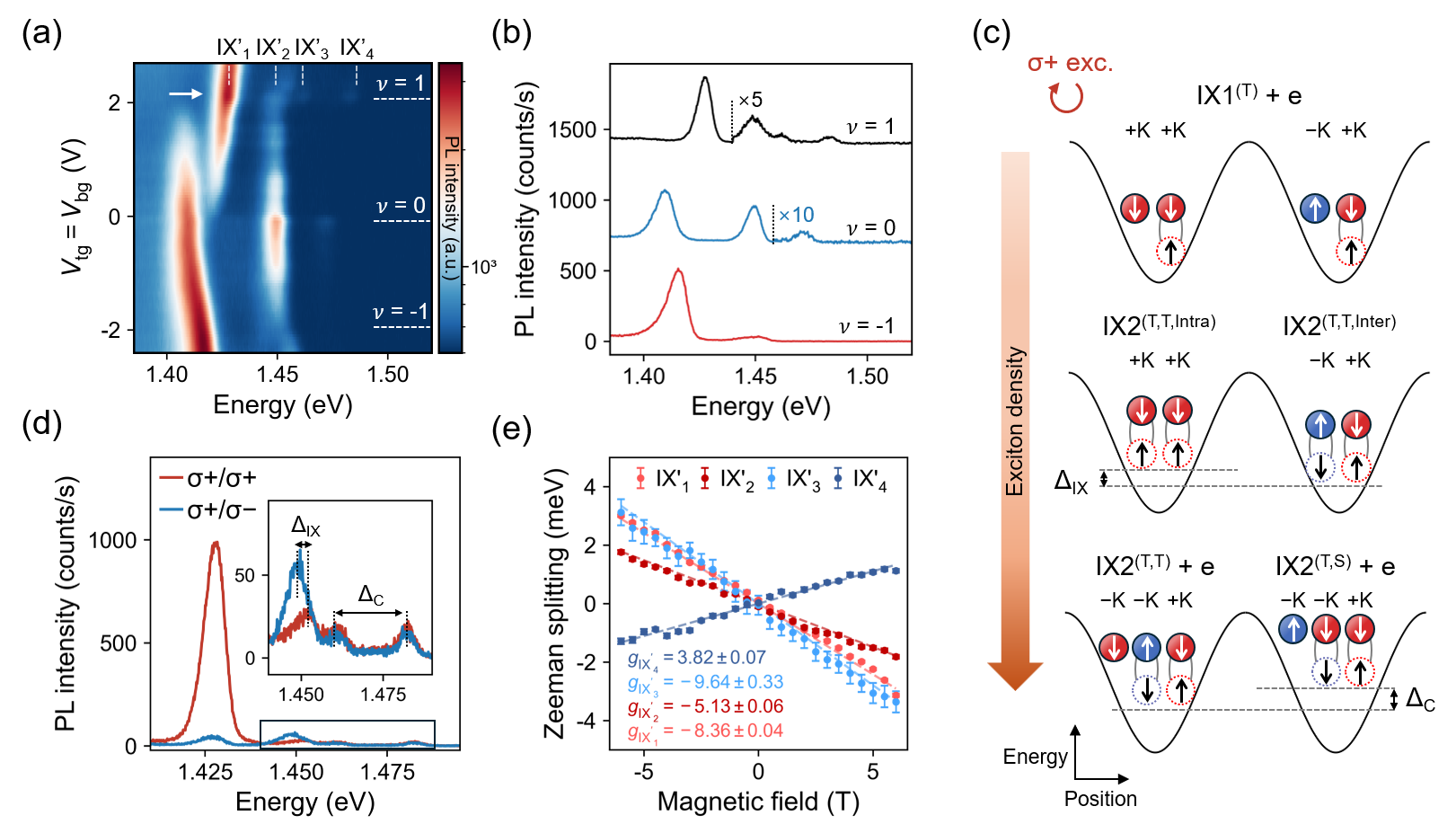}
    \caption{
    (a) Filling-dependent PL spectra measured at 7~$\mu$W under linearly polarized 1.680~eV excitation, showing the evolution of the IX emission from hole to electron filling.
    Additional resonances IX$'_1$--IX$'_4$ emerge prominently near $\nu=1$.
    (b) Representative linecuts at $\nu=-1$, 0, and 1.
    The high-energy spectral region is multiplied for clarity.
    (c) Schematic potential-energy landscape of the IX population dynamics under $\sigma^+$ excitation at $\nu=1$.
    (d) Helicity-resolved PL spectra at $\nu=1$ under $\sigma^+$ excitation.
    (e) Zeeman splittings of IX$'_1$--IX$'_4$ as a function of out-of-plane magnetic field.
    }
    \label{fig:fig4}
\end{figure*}

Having identified the spin--valley configurations of the charge-neutral exciton ladder, we next examine how resident carriers reshape the dipolar exciton spectrum.
Figure~4(a) shows filling-dependent PL spectra measured at 7~$\mu$W over the range from one hole per moir\'e site ($\nu=-1$) to one electron per moir\'e site ($\nu=1$) (see Supplemental Material, Fig.~S9 for the full filling-dependent spectra).
Representative linecuts at $\nu=-1$, 0, and 1 are shown in Fig.~4(b), displaying the asymmetric evolution of the IX ladder under hole and electron doping.
On the hole-doped side, IX$_1$ weakly blueshifts by approximately 6~meV at $\nu=-1$, while IX$_2$ gradually loses oscillator strength and separates into two peaks with distinct energies beyond $\nu=-1$ (see Supplemental Material, Sec.~S7 for a detailed discussion of the hole-doped regime).
We focus below on the electron-doped regime, where additional resonances emerge most prominently near $\nu=1$.

As the moir\'e lattice is populated with electrons, a new emission peak, IX$'_1$, emerges approximately 13~meV above IX$_1$.
We assign IX$'_1$ to a charged exciton--carrier complex, IX1$^{(\mathrm{T})}+e$, formed by the interaction between a moir\'e-trapped triplet IX and a resident electron occupying the same moir\'e site (Fig.~4(c))~\cite{miao2021strong-d88,park2023dipole-b29,kim2024correlation-driven-634}.
The intensity of IX$'_1$ increases as $\nu$ approaches 1, consistent with the growing population of localized electron--exciton complexes.
The behavior of IX$_2$ is markedly different.
Although its intensity is strongly modulated by electron filling, its emission energy remains close to that of the charge-neutral IX$_2$ even at $\nu=1$.
We therefore denote this resonance as IX$'_2$.
By contrast, IX$_3$ rapidly loses oscillator strength upon both electron and hole doping, indicating that the singlet-derived two-IX channel is strongly suppressed by the presence of resident carriers.
At $\nu=1$, the PL spectrum consists of IX$'_1$, IX$'_2$, and two additional resonances at 1.461~eV and 1.483~eV, denoted IX$'_3$ and IX$'_4$, respectively.
These two peaks are observed prominently only near $\nu=1$, suggesting that they are closely tied to the correlated electronic background established at one electron per moir\'e site.

Figure~4(d) shows helicity-resolved PL spectra at $\nu=1$ under $\sigma^+$ excitation. 
Similar to IX$_1$ at charge neutrality, IX$'_1$ exhibits strongly co-circularly polarized emission, with $\rho$ approaching 0.9. 
This enhanced polarization of IX$'_1$ compared with charge-neutral IX$_1$ ($\rho\sim0.7$) may arise from the shorter lifetime of the charged excitonic complex, which suppresses exchange-mediated valley depolarization~\cite{wu2025highly-80a,jin2018ultrafast-bb4}.
Strikingly, IX$'_2$ exhibits strong cross-circular polarization, in contrast to charge-neutral IX$_2$, which is nearly unpolarized (Fig.~3(b)). 
Moreover, the co-circular component of IX$'_2$ appears at a slightly higher energy than the cross-circular component, with an energy splitting of $\sim$2.8~meV (inset of Fig.~4(d)). 
Neither the $\rho$ nor the energy splitting is observed under linearly polarized excitation, indicating that both originate from optically induced valley-selective populations (see Supplemental Material, Fig.~S10).

The proximity of IX$'_2$ to the charge-neutral IX$_2$ resonance suggests that it retains a neutral two-exciton-like character, rather than forming a completely distinct charged excitonic complex.
We therefore attribute the lower- and higher-energy components of IX$'_2$ to intervalley and intravalley IX2$^{(\mathrm{T,T})}$ configurations, respectively, as illustrated in Fig.~4(c). 
Under $\sigma^+$ excitation resonant with the WSe$_2$ 1$s$ exciton, $K$-valley IX$^{(\mathrm{T})}$ states are preferentially generated. 
Because the electronic background at $\nu=1$ is not known to be valley polarized in WSe$_2$/WS$_2$~\cite{tang2020simulation-64e,park2023dipole-b29}, the preexisting electrons may occupy either the $+K$ or $-K$ valley. 
When a resident electron participates in the formation of a second IX$^{(\mathrm{T})}$ in the same valley as the optically generated IX, an intravalley ($+K,+K$) two-exciton configuration is realized.
By contrast, participation of an electron from the opposite valley leads to an intervalley two-exciton configuration ($+K,-K$).
The lower energy of the cross-circular component is consistent with the intervalley configuration being energetically favored over the intravalley one. 
The stronger cross-circular polarization of IX$'_2$ further suggests that this lower-energy intervalley channel is preferentially populated under circularly polarized excitation.
The observed splitting is comparable to the intravalley--intervalley two-exciton splitting reported in H-stacked WSe$_2$/WS$_2$~\cite{jiang2025tuning-dc4}, supporting our assignment to intervalley and intravalley two-IX configurations.

We next consider the origin of IX$'_3$ and IX$'_4$. 
Both peaks are nearly unpolarized, similar to IX$_2$ and IX$_3$ at charge neutrality. 
One possible interpretation is that they originate from moir\'e trapping sites with a different atomic registry. 
This scenario is unlikely, however, since the calculated energy offset between the $R_{\mathrm{h}}^{X}$ site and the next local minimum is expected to be on the order of $\sim$100~meV or larger~\cite{zhu2024moir-b7a,yuan2020twist-angle-dependent-439}, whereas IX$'_3$ and IX$'_4$ lie only 52~meV and 74~meV above IX$_1$, respectively.
Consistent with this picture, the out-of-plane electric-field dependence of IX$'_1$--IX$'_4$ yields similar electron--hole separations of $\sim$0.7~nm from the measured DC Stark shifts, indicating that all four peaks originate from the moir\'e sites with the same local atomic registry (Fig.~S11).
The excitation-power-dependent PL intensity further supports a many-body origin for IX$'_3$ and IX$'_4$.
Both IX$'_3$ and IX$'_4$ emerge at higher onset powers than IX$'_1$ and IX$'_2$ and exhibit strongly nonlinear power-law scaling, with exponents of $\alpha \sim 3.4$ and $4.0$, respectively (Fig.~S12).
These large exponents suggest that IX$'_3$ and IX$'_4$ emerge through higher-order formation pathways in the correlated electron--IX background, following the sequential occupation of lower-energy charged IX states.

The energy hierarchy closely mirrors that of the charge-neutral IX ladder.
The energy separations between IX$'_1$ and IX$'_3$, and between IX$'_3$ and IX$'_4$, are approximately 34~meV and 22~meV, respectively, comparable to the corresponding energy separations between IX$_1$, IX$_2$, and IX$_3$ at $\nu=0$. 
Furthermore, helicity-resolved magneto-PL measurements reveal distinct Zeeman responses for IX$'_3$ and IX$'_4$ (Fig.~4(e)). 
The extracted $g$-factors are $g_{\mathrm{IX}'_3}=-9.64\pm0.33$ and $g_{\mathrm{IX}'_4}=3.82\pm0.07$, respectively, closely matching those of IX$2^{(\mathrm{T,T})}$ and IX$2^{(\mathrm{T,S})}$ at charge neutrality.
The modest renormalization of $g$-factors relative to the neutral case may arise from exchange interactions within the electron--exciton complex~\cite{campbell2022exciton-polarons-384}. 
Taken together, these observations identify IX$'_3$ and IX$'_4$ as charged two-exciton states, IX$2^{(\mathrm{T,T})}+e$ and IX$2^{(\mathrm{T,S})}+e$, respectively (Fig.~4(c)). 
Their emergence only in the vicinity of $\nu=1$ indicates that they originate from interactions between multiple IXs and a localized electron embedded in the correlated electronic background \cite{regan2020mott-1b0}. 
Thus, the resident electron acts as a localized fermionic degree of freedom coupled to the bosonic dipolar ladder.
We finally note that the energy separation between IX$'_1$ and IX$'_3$ is around 34~meV, slightly smaller than the charge-neutral on-site dipolar repulsion $U_{dd}\sim$38~meV.
The reduction indicates that the resident electron partially screens the IX--IX interaction, leading to a renormalization of the effective dipolar repulsion within the moiré lattice.

In summary, we identify the spin--valley configurations of multiple moir\'e-confined IX states in R-stacked WSe$_2$/WS$_2$ heterobilayers.
Polarization-resolved magneto-PL measurements reveal that the observed peaks do not form a simple dipolar ladder generated solely by the successive occupation of the ground-state IX$^{(\mathrm{T})}$ excitons. 
Instead, the spin-split WS$_2$ conduction band gives rise to distinct triplet--triplet and triplet--singlet two-exciton configurations, establishing the microscopic origin of the unequal ladder spacing at charge neutrality.
At electron filling $\nu=1$, the correlated electronic background reshapes the emission spectrum, giving rise to charged electron--IX complexes and reducing the effective interaction energy of the dipolar ladder. 
We further observe a helicity-dependent fine structure of the two-IX emission, consistent with the coexistence of intervalley and intravalley configurations. 
Our results establish TMD heterobilayers as a promising platform for exploring spin-dependent Bose--Fermi mixtures, where the spin--valley degrees of freedom of excitons and resident carriers jointly determine the formation of correlated excitonic complexes.

\begin{acknowledgments}
This work was supported by the EPSRC (grant nos. EP/P029892/1 and EP/Y026284/1).
B. W. C. is supported by a Marie Skłodowska-Curie Individual Fellowship (No. 101208787).
M.B.-G. is supported by a Royal Society University Research Fellowship. B.D.G. is supported by a Chair in Emerging Technology from the Royal Academy of Engineering. 
K.W. and T.T. acknowledge the support from the JSPS KAKENHI (Grant Numbers 21H05233 and 23H02052) , the CREST (JPMJCR24A5), JST and World Premier International Research Center Initiative (WPI), MEXT, Japan.
\end{acknowledgments}

\bibliography{reference}

\end{document}


\title{Supplemental Material for\\
``Unveiling the Spin--Valley Structure of Dipolar Exciton Ladders in R-stacked WSe$_2$/WS$_2$ Moir\'e Heterobilayers''}

\author{Byeong Wook Cho}
\thanks{These authors contributed equally to this work.}
\affiliation{Institute of Photonics and Quantum Sciences, SUPA, Heriot-Watt University, Edinburgh, UK}

\author{Tatyana V. Ivanova}
\thanks{These authors contributed equally to this work.}
\affiliation{Institute of Photonics and Quantum Sciences, SUPA, Heriot-Watt University, Edinburgh, UK}

\author{Zhe Li}
\affiliation{Institute of Photonics and Quantum Sciences, SUPA, Heriot-Watt University, Edinburgh, UK}

\author{Takashi Taniguchi}
\affiliation{Research Center for Materials Nanoarchitectonics, National Institute for Materials Science, 1-1 Namiki, Tsukuba 305-0044, Japan}

\author{Kenji Watanabe}
\affiliation{Research Center for Electronic and Optical Materials, National Institute for Materials Science, 1-1 Namiki, Tsukuba 305-0044, Japan}

\author{Brian D. Gerardot}
\affiliation{Institute of Photonics and Quantum Sciences, SUPA, Heriot-Watt University, Edinburgh, UK}

\author{Mauro Brotons-Gisbert}
\affiliation{Institute of Photonics and Quantum Sciences, SUPA, Heriot-Watt University, Edinburgh, UK}

\date{\today}

\maketitle

\section{S1. Device fabrication and twist-angle determination}
The R-stacked WSe$_2$/WS$_2$ moir\'e heterobilayer devices were fabricated using a dry-transfer pickup technique.
Monolayer WSe$_2$, monolayer WS$_2$, few-layer hBN, and graphite flakes were mechanically exfoliated from bulk crystals and identified by optical contrast.
The monolayer nature of the TMD flakes was further confirmed by photoluminescence (PL) measurements.
The heterostructures were assembled using a polycarbonate (PC)/PDMS stamp.
The WSe$_2$ and WS$_2$ monolayers were sequentially picked up together with hBN flakes while controlling their relative orientation to obtain R-type stacking with small twist angles.
The completed stack was released onto a prepatterned electrode substrate by heating the PC film at $200^\circ$C, followed by removal of residual PC in chloroform.

The crystallographic orientations of the WSe$_2$ and WS$_2$ monolayers were first estimated from straight crystal edges and then verified by polarization-resolved second-harmonic generation (SHG) measurements.
Figure~S1 shows optical microscope images of devices R1--R3 together with the corresponding SHG polar plots measured from the WSe$_2$ and WS$_2$ monolayers.

\begin{figure}[t]
    \centering
    \includegraphics[width=\linewidth]{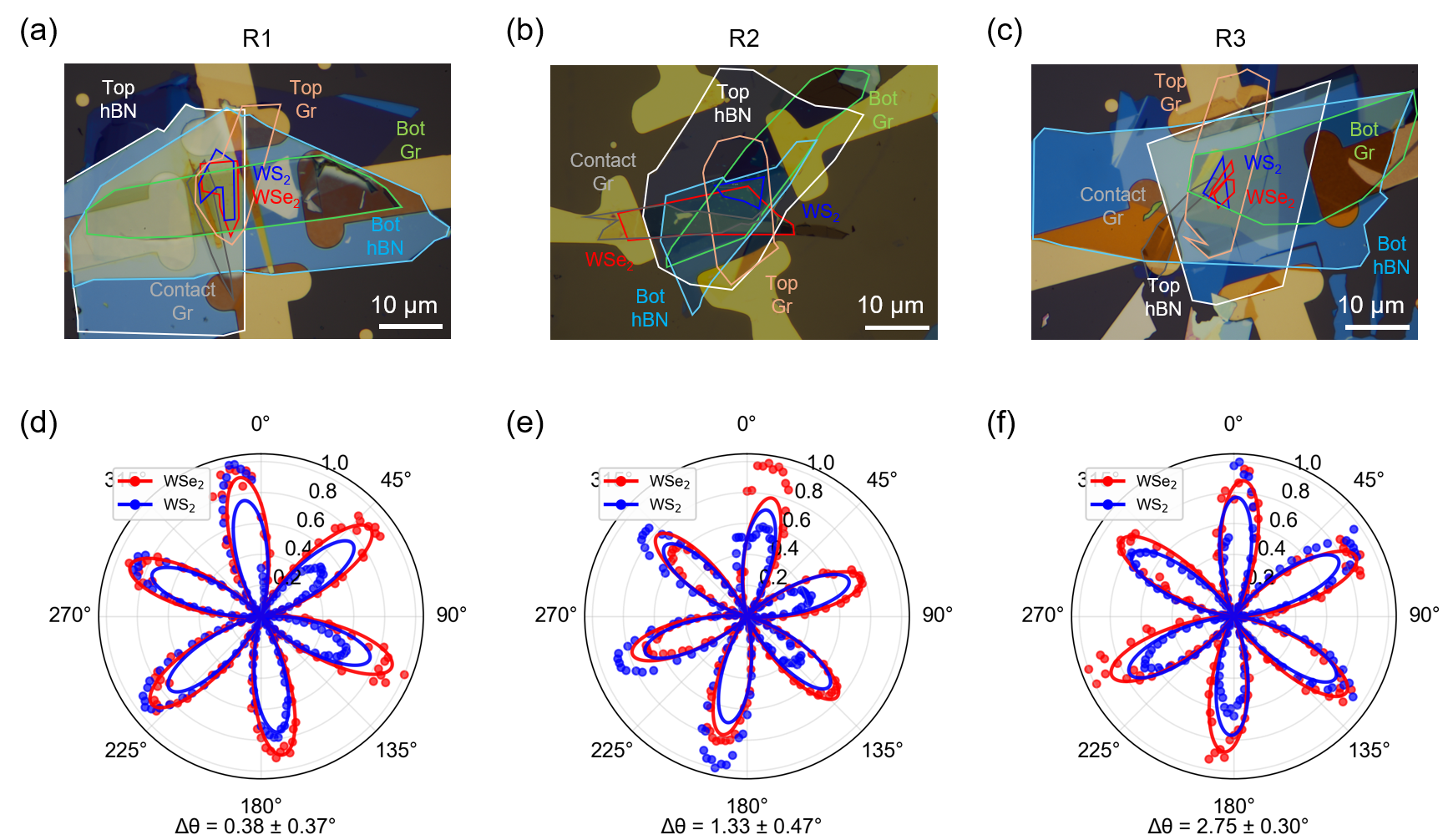}
    \caption{
    Device images and polarization-resolved SHG characterization of R-stacked WSe$_2$/WS$_2$ moir\'e heterobilayers.
    (a)--(c) Optical microscope images of devices R1--R3.
    (d)--(f) Corresponding SHG polar plots of the WSe$_2$ and WS$_2$ monolayers.
    The extracted twist angles are $0.38 \pm 0.37^\circ$, $1.33 \pm 0.47^\circ$, and $2.75 \pm 0.30^\circ$ for R1, R2, and R3, respectively.
    }
    \label{fig:S1}
\end{figure}

\begin{figure}[t]
    \centering
    \includegraphics[width=0.5\linewidth]{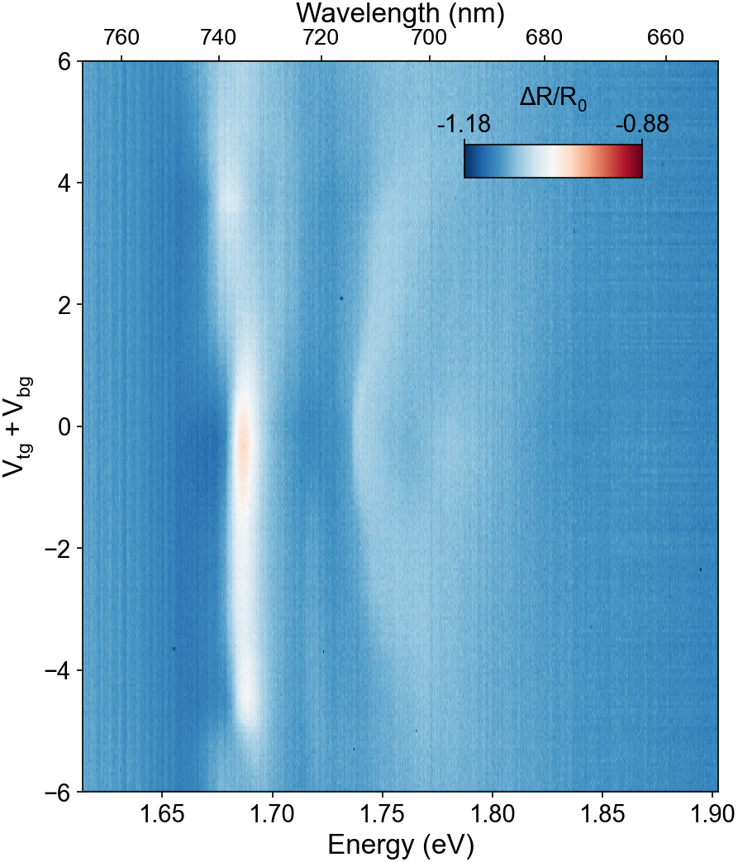}
    \caption{
    Reflectance contrast of device R1 near the WSe$_2$ intralayer exciton resonance.
    The main WSe$_2$ 1$s$ resonance appears at approximately 1.680~eV, accompanied by higher-energy moir\'e-induced spectral features.
    }
    \label{fig:S2}
\end{figure}

\clearpage
\section{S2. Capacitance model}
The carrier density and out-of-plane electric field in the dual-gated WSe$_2$/WS$_2$ moir\'e heterobilayer devices were estimated using a parallel-plate capacitance model.
The top and bottom graphite gates are separated from the WSe$_2$/WS$_2$ heterobilayer by hBN dielectric layers with thicknesses $d_\mathrm{t}$ and $d_\mathrm{b}$, respectively.
For device R1, both the top and bottom hBN thicknesses are approximately 40~nm.
Taking the gate voltages relative to the charge-neutrality point, the induced carrier density is given by
\begin{equation}
n =
\frac{\epsilon_0 \epsilon_{\mathrm{hBN}}}{e}
\left(
\frac{V_\mathrm{tg}}{d_\mathrm{t}}
+
\frac{V_\mathrm{bg}}{d_\mathrm{b}}
\right),
\end{equation}
where $\epsilon_0 = 8.854 \times 10^{-12}$~F/m is the vacuum permittivity, $\epsilon_{\mathrm{hBN}} \approx 3.8$ is the relative dielectric constant of hBN, and $e$ is the elementary charge.

The out-of-plane electric field is estimated from the difference between the top- and bottom-gate displacement fields:
\begin{equation}
E_z =
\frac{\epsilon_{\mathrm{hBN}}}{2}
\left(
\frac{V_\mathrm{tg}}{d_\mathrm{t}}
-
\frac{V_\mathrm{bg}}{d_\mathrm{b}}
\right).
\end{equation}
This dual-gate geometry allows the carrier density and vertical electric field to be tuned independently by sweeping $V_\mathrm{tg}$ and $V_\mathrm{bg}$ along appropriate trajectories in gate-voltage space.

\clearpage
\section{S3. Optical selection rule}
In R-stacked WSe$_2$/WS$_2$ heterobilayers, the $K$ valleys of WSe$_2$ and WS$_2$ are nearly momentum-aligned, as are the corresponding $-K$ valleys.
The moir\'e-trapped IXs considered here are localized at the $R_{\mathrm{h}}^{X}$ site, so their emission helicity is determined by the spin--valley indices of the electron and hole involved in recombination.
The optical selection rules for interband transitions at the relevant local registries are summarized in Table~S1.

Following the convention in Table~S1, the $+K$ and $-K$ valley triplet transitions at the $R_{\mathrm{h}}^{X}$ site couple predominantly to $\sigma^+$ and $\sigma^-$ polarized light, respectively, whereas the corresponding singlet transitions couple to the opposite helicities, $\sigma^-$ and $\sigma^+$, respectively.
These selection rules provide the basis for assigning the spin--valley character of the IX peaks in the main text.

\begin{table}[h!]
\centering
\caption{
Optical selection rules for interband transitions at different local registries in R-stacked WSe$_2$/WS$_2$, adapted from \cite{wang2023intercell-416}
}
\setlength{\tabcolsep}{15pt}
\begin{tabular}{c|ccc}
\hline
Transition & $R_{\mathrm{h}}^{h}$ & $R_{\mathrm{h}}^{M}$ & $R_{\mathrm{h}}^{X}$ \\
\hline
$|K_{\mathrm{v}}\uparrow\rangle \rightarrow |K'_{\mathrm{c}}\uparrow\rangle$ 
& $\sigma_+$ & $z$ & $\sigma_-$ \\
$|K_{\mathrm{v}}\uparrow\rangle \rightarrow |K'_{\mathrm{c}}\downarrow\rangle$ 
& $z$ & $\sigma_-$ & $\sigma_+$ \\
$|-K_{\mathrm{v}}\downarrow\rangle \rightarrow |-K'_{\mathrm{c}}\downarrow\rangle$ 
& $\sigma_-$ & $z$ & $\sigma_+$ \\
$|-K_{\mathrm{v}}\downarrow\rangle \rightarrow |-K'_{\mathrm{c}}\uparrow\rangle$ 
& $z$ & $\sigma_+$ & $\sigma_-$ \\
\hline
\end{tabular}
\label{tab:selection_rules}
\end{table}

\clearpage
\begin{figure}[t]
    \centering
    \includegraphics[width=0.8\linewidth]{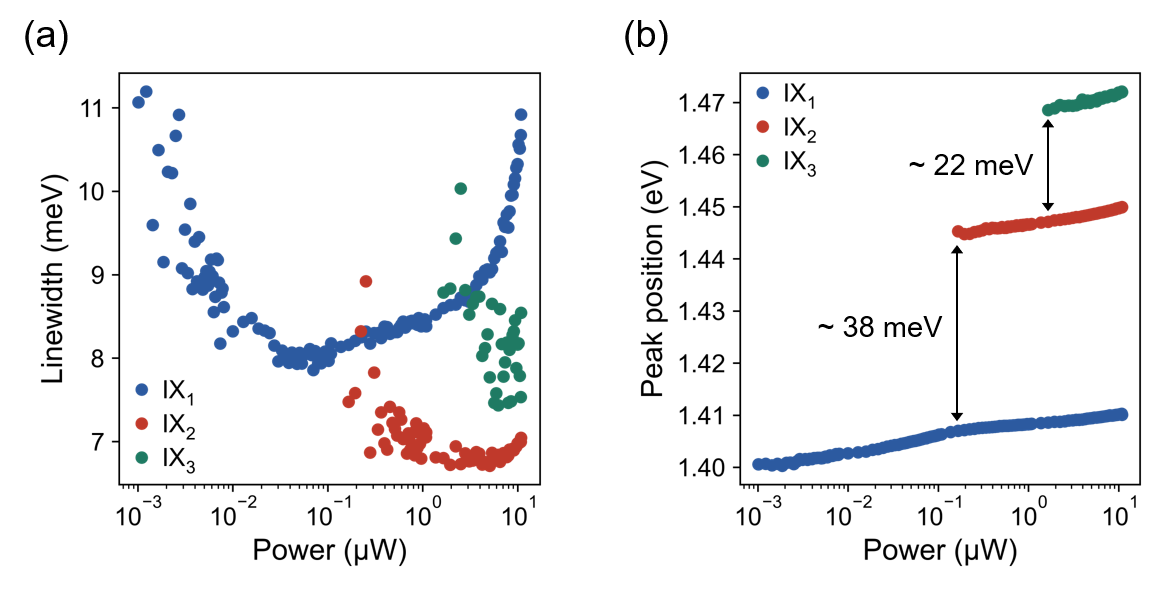}
    \caption{
    Power-dependent PL peak parameters of device R1 at charge neutrality.
    (a) Extracted linewidths and (b) peak energies of IX$_1$, IX$_2$, and IX$_3$ as a function of excitation power.
    }
    \label{fig:S3}
\end{figure}

\clearpage
\begin{figure}[t]
    \centering
    \includegraphics[width=\linewidth]{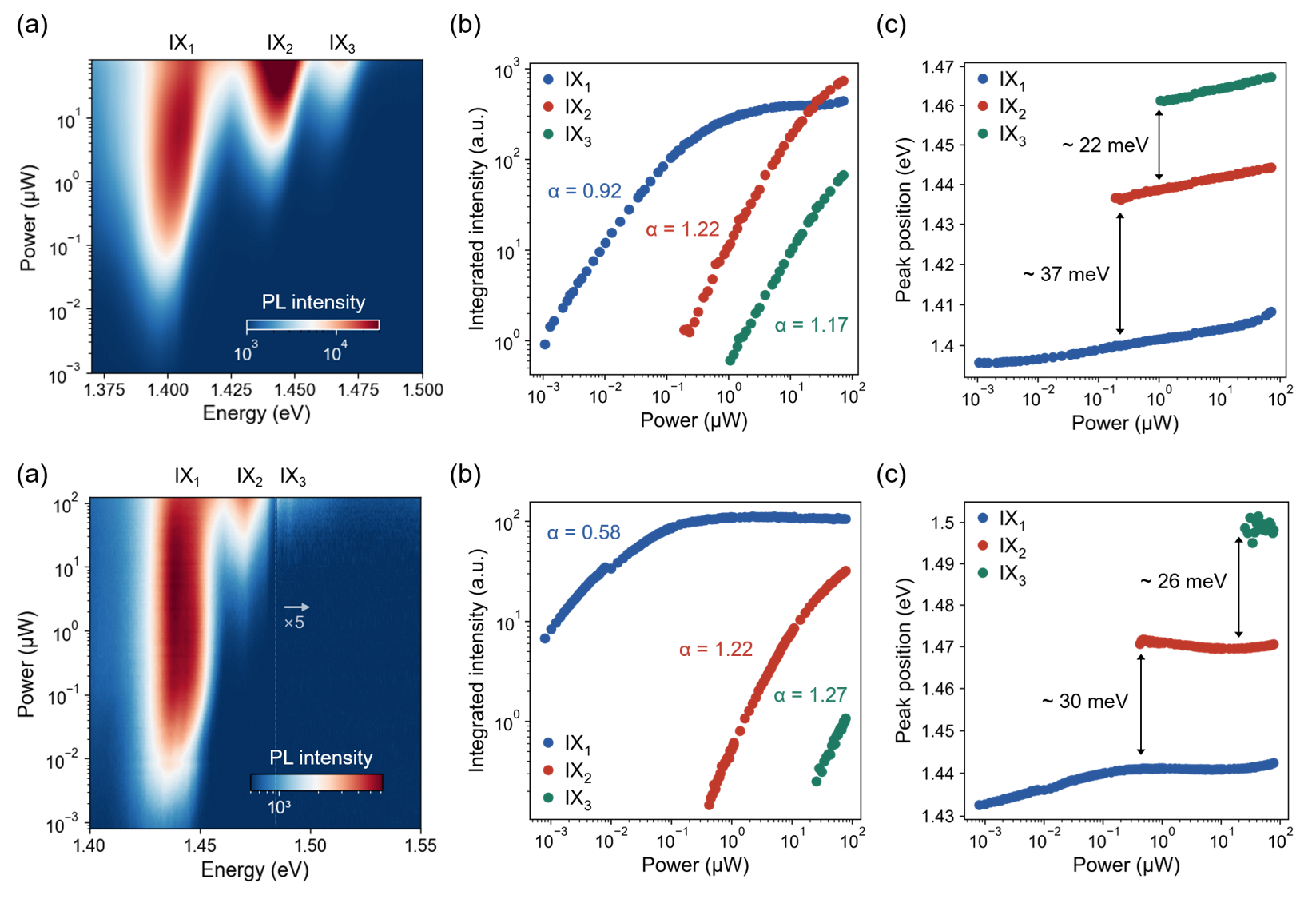}
    \caption{
    Power-dependent PL measurements of additional R-stacked WSe$_2$/WS$_2$ moir\'e heterobilayer devices.
    (a)--(c) Data from device R2: power-dependent PL map, integrated PL intensities, and peak energies of IX$_1$, IX$_2$, and IX$_3$.
    (d)--(f) Corresponding data from device R3.
    The emergence of higher-energy IX$_2$ and IX$_3$ peaks at elevated excitation powers, together with their nonlinear power dependence, is consistent with the behavior observed in device R1.
    }
    \label{fig:S4}
\end{figure}

\begin{figure}[t]
    \centering
    \includegraphics[width=0.8\linewidth]{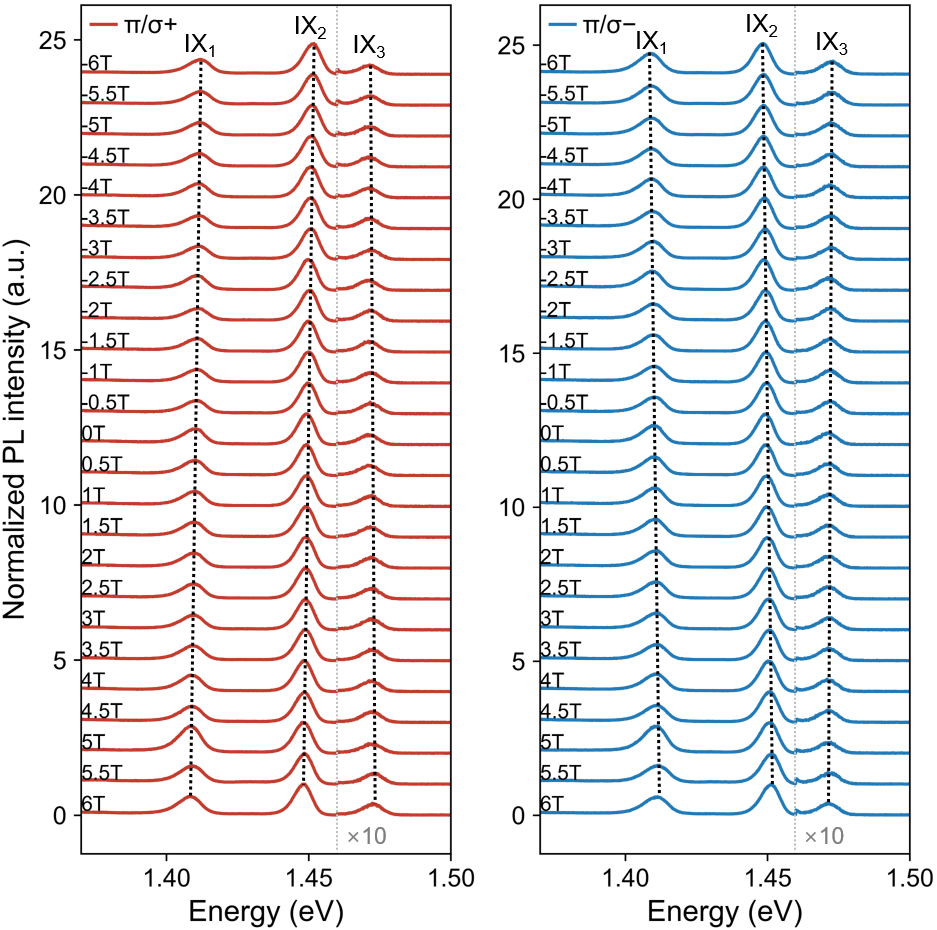}
    \caption{
    Helicity-resolved magneto-PL spectra of device R1 at charge neutrality under linearly polarized 1.72~eV excitation.
    (a) PL spectra detected in the $\sigma^+$ channel.
    (b) PL spectra detected in the $\sigma^-$ channel.
    The out-of-plane magnetic field was swept from $-6$~T to $+6$~T in steps of 0.5~T.
    The IX$_3$ emission is magnified by a factor of 10 for clarity.
    }
    \label{fig:S5}
\end{figure}

\clearpage
\section{S4. \texorpdfstring{$g$}{g}-factor calculation}
The effective exciton $g$-factors were extracted from helicity-resolved magneto-PL measurements (see Fig.~S5). 
For each magnetic field, the PL spectra collected in the $\sigma^+$ and $\sigma^-$ detection channels were fitted with Gaussian functions to determine the peak energies $E_{\sigma^+}(B)$ and $E_{\sigma^-}(B)$.
The Zeeman splitting was then defined as
\begin{equation}
\Delta E_Z(B) = E_{\sigma^+}(B) - E_{\sigma^-}(B).
\end{equation}

The magnetic-field dependence of the Zeeman splitting was fitted using
\begin{equation}
\Delta E_Z(B) = g \mu_B B,
\end{equation}
where $\mu_B = 57.88~\mu\mathrm{eV/T}$ is the Bohr magneton and $g$ is the effective exciton $g$-factor.
The $g$-factor was obtained from the slope of the linear fit,
\begin{equation}
g = \frac{1}{\mu_B}\frac{d\Delta E_Z}{dB}.
\end{equation}
The sign of $g$ is determined by the relative energy shift of the $\sigma^+$ and $\sigma^-$ emission components with magnetic field.
Positive and negative values therefore indicate opposite valley-Zeeman responses of the corresponding IX states.

\begin{figure}[t]
    \centering
    \includegraphics[width=\linewidth]{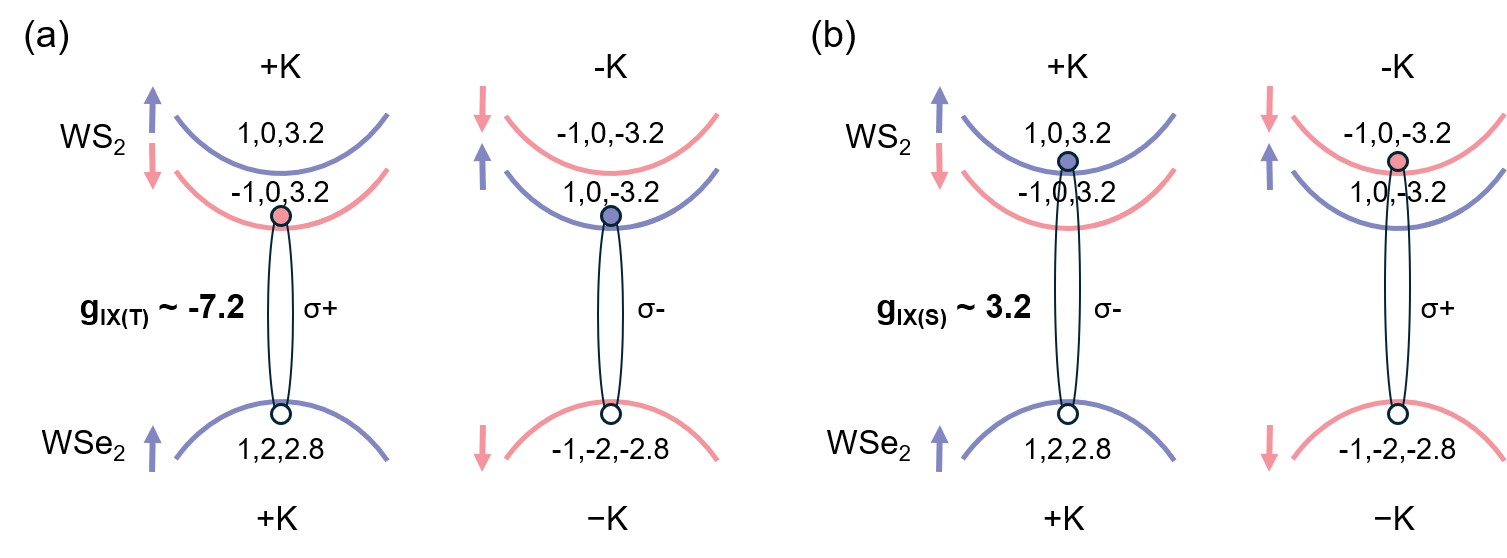}
    \caption{
    Schematic illustration of the contributions to the valley-Zeeman response of interlayer exciton transitions.
    (a) Spin-triplet transition and (b) spin-singlet transition considered for the effective $g$-factor estimation.
    The magnetic-field-induced energy shift of the conduction or valence band edge is decomposed as
    $\Delta E_{c/v}=\Delta_s+\Delta_a+\Delta_v$, where $\Delta_s$ is the spin contribution, $\Delta_a$ is the atomic orbital contribution, and $\Delta_v$ is the valley magnetic moment contribution associated with the Berry curvature.
    The numerical labels indicate the representative spin, atomic orbital, and valley contributions used to estimate the effective Zeeman response of each transition.
    The Berry-curvature-related valley contributions were taken from~\cite{kormnyos2015corrigendum-e8f}.
    }
    \label{fig:S6}
\end{figure}

\clearpage
\section{S5. Polarization calibration}
To obtain reliable degrees of circular polarization, we calibrated the polarization-dependent response of the optical setup. 
In polarization-resolved PL measurements, the measured intensity can be affected not only by the sample emission but also by helicity-dependent transmission or detection efficiency of the optical path. 
This artificial asymmetry can arise from polarization-dependent reflection and transmission of optical components, most likely the beam splitter in our setup.

The calibration procedure is illustrated in Fig.~S7. 
A reference laser signal was sent through the collection path, and the detected intensity was recorded while rotating the half-wave plate in the collection arm. 
In an ideal polarization-independent detection path, the measured intensity should remain constant as the analyzer angle is varied. 
Instead, we observed a weak angular modulation of the detected intensity, indicating a finite polarization-dependent response of the setup.

The measured modulation curve was used as a calibration factor for the helicity-resolved PL spectra. 
The raw $\sigma^+$ and $\sigma^-$ intensities were corrected by the corresponding detection efficiencies before calculating the degree of circular polarization,
\begin{equation}
\rho =
\frac{I_{\sigma^+}-I_{\sigma^-}}
{I_{\sigma^+}+I_{\sigma^-}}.
\end{equation}
This procedure removes setup-induced polarization asymmetry and allows the extracted polarization values to reflect the intrinsic helicity response of the IX emission.

\clearpage
\begin{figure}[t]
    \centering
    \includegraphics[width=\linewidth]{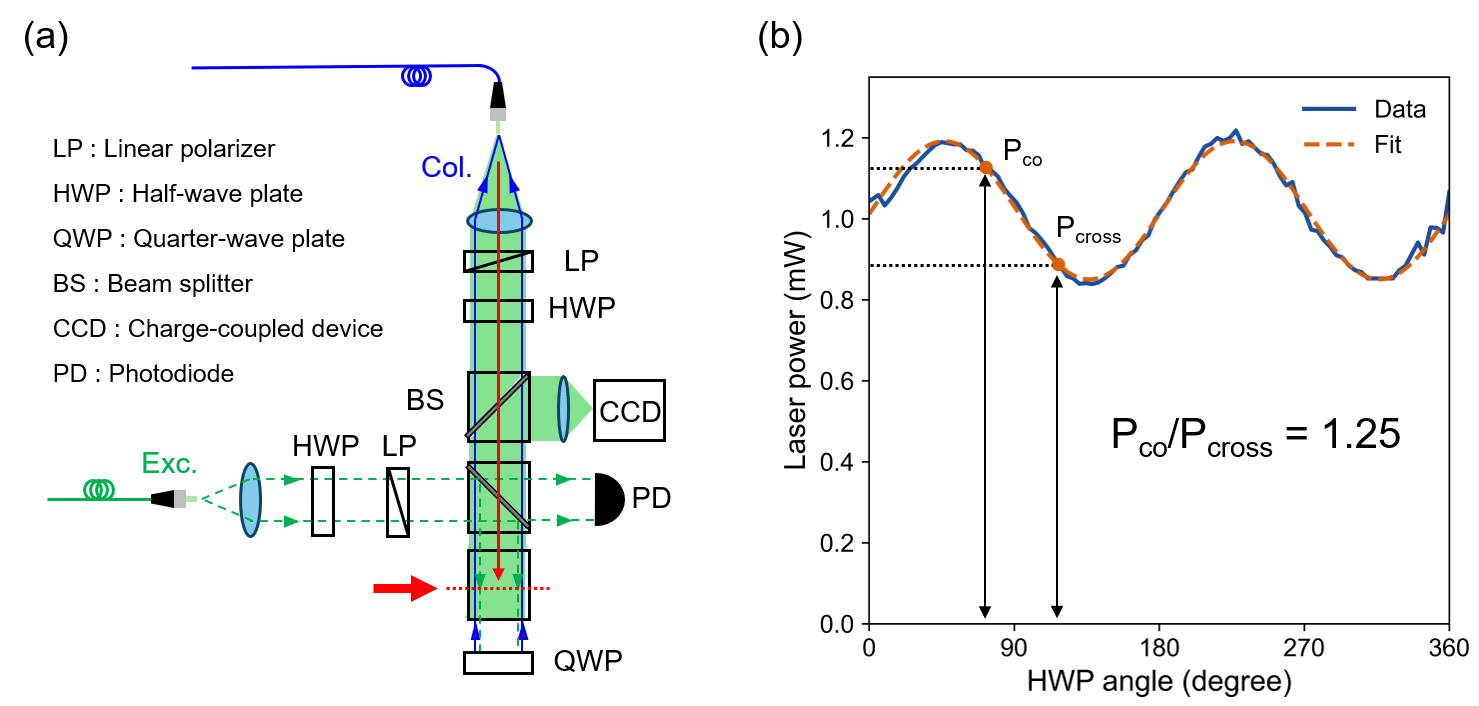}
    \caption{
    Optical setup and polarization calibration.
    (a) Schematic of the excitation and collection paths used for helicity-resolved PL measurements, showing the polarization-control and polarization-analysis optics.
    A reference laser signal was sent through the collection path, and the detected intensity was recorded while rotating the half-wave plate in the collection arm.
    (b) Measured angular dependence of the reference signal, together with a sinusoidal fit.
    The observed modulation reflects the polarization-dependent response of the optical setup, mainly arising from the beam splitters.
    This calibration curve was used to correct the raw $\sigma^+$ and $\sigma^-$ PL intensities.
    }
    \label{fig:S7}
\end{figure}

\clearpage
\section{S6. Exchange interaction}
The magnetic-field dependence of the degree of circular polarization provides an estimate of the exchange-interaction strength~\cite{she2025magneto-polarization-6b6,wu2025highly-80a,jiang2025tuning-dc4}.
The first panel of Fig.~3(d) in the main text and the power-dependent data shown in Fig.~S8 were fitted using the following model:
\begin{equation}
\rho(B)=
\frac{\rho_0}
{1+\dfrac{2\tau/\tau_{v0}}{1+\left(B/B_c\right)^2}},
\label{eq:exchange_rho_model}
\end{equation}
where $\rho_0$ is the initial degree of circular polarization in the absence of valley depolarization, $\tau_{v0}$ is the valley depolarization time at zero magnetic field, and $\tau$ is the exciton lifetime.
The characteristic magnetic field $B_c$ represents the field scale at which the out-of-plane magnetic field suppresses exchange-driven valley depolarization, and therefore provides an estimate of the exchange-interaction strength.

Following previous reports~\cite{wu2025highly-80a,jiang2025tuning-dc4}, we estimate the corresponding exchange-energy scale as
\begin{equation}
\Delta_{\mathrm{ex}} \sim |g|\mu_B B_c ,
\label{eq:exchange_energy}
\end{equation}
where $\mu_B=57.88~\mu\mathrm{eV/T}$ is the Bohr magneton and $g$ is the effective exciton $g$-factor.
Using the measured value $|g|=8.73$ for IX$_1$, the extracted characteristic fields $B_c=0.05$, 0.04, and 0.03~T at excitation powers of 30, 150, and 500~nW correspond to $\Delta_{\mathrm{ex}}\approx0.025$, 0.020, and 0.015~meV, respectively.

\clearpage

\begin{figure}[t]
    \centering
    \includegraphics[width=0.7\linewidth]{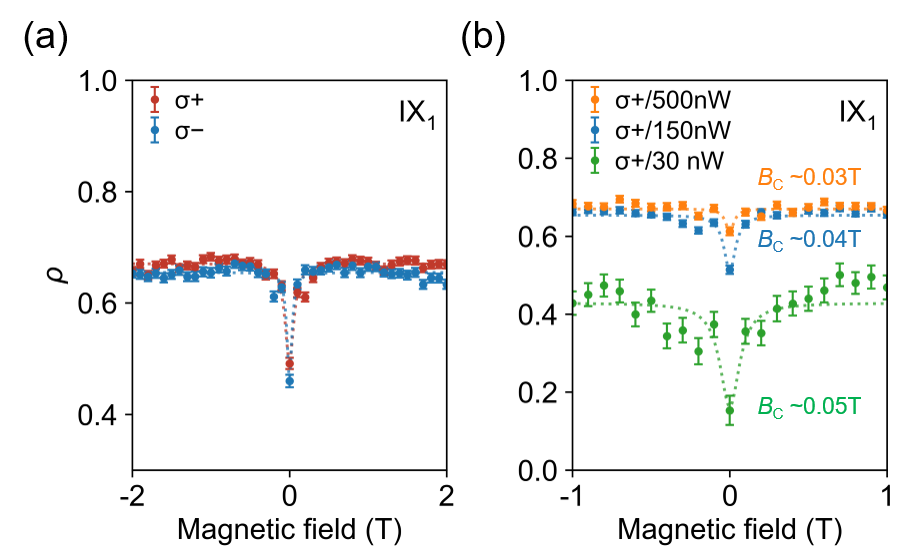}
    \caption{
    Magnetic-field dependence of the degree of circular polarization of IX$_1$.
    (a) Full-range polarization data under $\sigma^+$ and $\sigma^-$ excitation, extending the field range shown in Fig.~3(d).
    (b) Power-dependent polarization data fitted to extract the characteristic field $B_{\mathrm{c}}$ for the rapid polarization increase.
    The extracted $B_{\mathrm{c}}$ values remain similar, $\sim$0.03--0.05~T, across the investigated excitation-power range, while a residual depolarization persists even at high magnetic fields.
    }
    \label{fig:S8}
\end{figure}

\begin{figure}[t]
    \centering
    \includegraphics[width=0.5\linewidth]{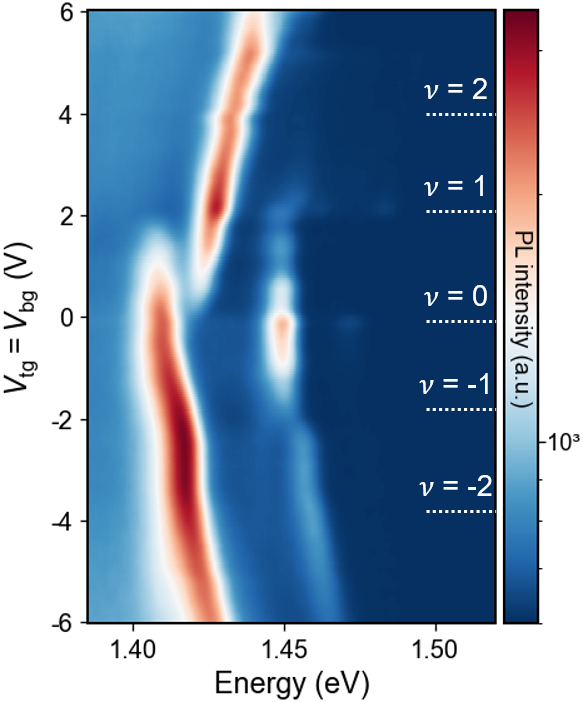}
    \caption{
    Full filling-dependent PL spectra of device R1 measured at an excitation power of 7~$\mu$W.
    Dashed lines indicate representative integer filling factors $\nu=-2$, $-1$, 0, 1, and 2.
    The zoomed-in range between $\nu=-1$ and $\nu=1$ is discussed in Fig.~4(a).
    }
    \label{fig:S9}
\end{figure}

\begin{figure}[t]
    \centering
    \includegraphics[width=0.6\linewidth]{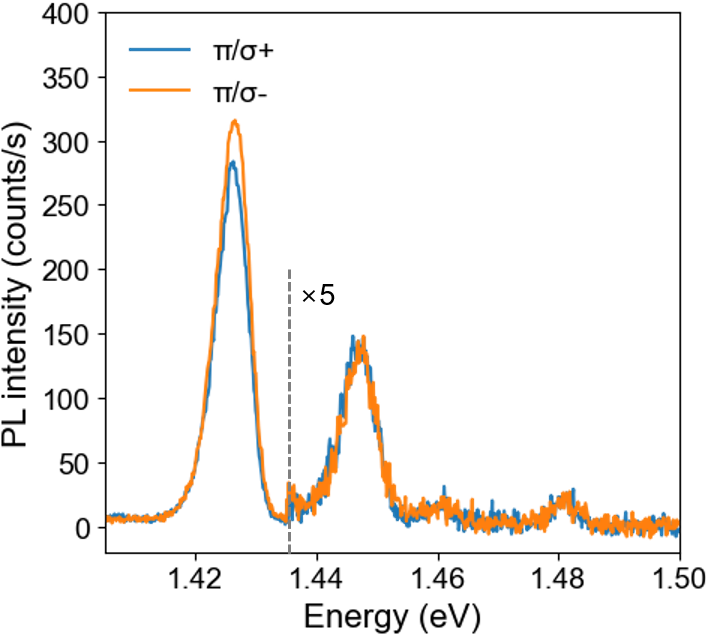}
    \caption{
    Helicity-resolved PL spectra of device R1 at $\nu=1$ under linearly polarized excitation.
    The $\sigma^+$ and $\sigma^-$ detection channels show no clear circular polarization or helicity-dependent splitting, confirming that the splitting of IX$'_2$ in Fig.~4(d) is induced by circularly polarized excitation.
    }
    \label{fig:S10}
\end{figure}

\clearpage
\begin{figure}[t]
    \centering
    \includegraphics[width=0.8\linewidth]{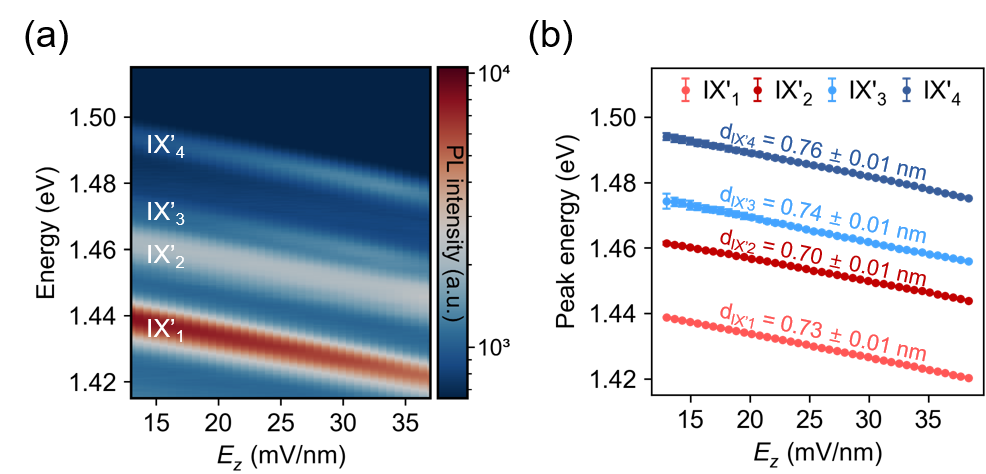}
    \caption{
    Out-of-plane electric-field dependence of IX$'_1$--IX$'_4$ at $\nu=1$.
    (a) PL spectra as a function of vertical electric field.
    (b) Extracted peak energies with linear Stark-shift fits.
    The fitted electron--hole separations are similar for all four peaks, $d\sim0.7$~nm, indicating that IX$'_1$--IX$'_4$ originate from the same moir\'e site.
    }
    \label{fig:S11}
\end{figure}

\clearpage
\begin{figure}[t]
    \centering
    \includegraphics[width=0.85\linewidth]{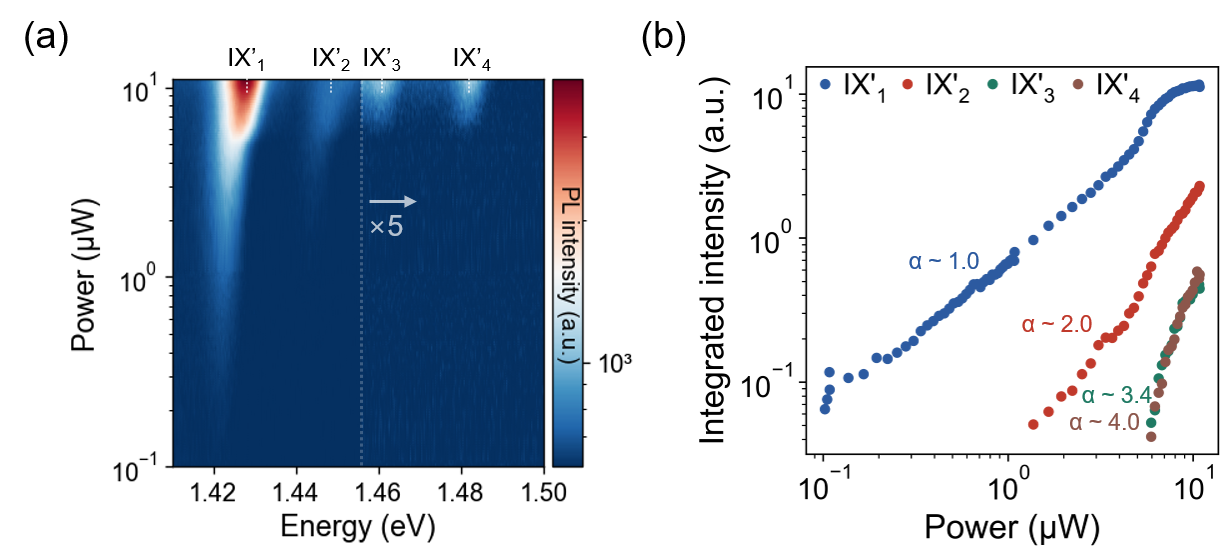}
    \caption{
    Excitation-power dependence of the electron-doped IX peaks at $\nu=1$.
    (a) Power-dependent PL map showing the emergence of IX$'_1$--IX$'_4$.
    The high-energy spectral region containing IX$'_3$ and IX$'_4$ is magnified by a factor of 5 for clarity.
    (b) Integrated PL intensities of IX$'_1$--IX$'_4$ as a function of excitation power, plotted on a log--log scale.
    The extracted power-law exponents are $\alpha\sim1.0$, 2.0, 3.4, and 4.0 for IX$'_1$, IX$'_2$, IX$'_3$, and IX$'_4$, respectively.
    }
    \label{fig:S12}
\end{figure}

\clearpage
\section{S7. Dipolar ladders in WSe$_2$/WS$_2$ at \texorpdfstring{$\nu=-1$}{nu = -1}}

We further examined the dipolar IX emission in the hole-doped regime.
Figure~S13(a) shows the PL spectrum at $\nu=-1$, where three emission features are observed and denoted IX$_1^{+}$, IX$_2^{+}$, and IX$_3^{+}$.
Here, the superscript $+$ indicates features observed in the hole-doped regime.

In contrast to the electron-doped side, where a distinct charged exciton--carrier complex IX$'_1$ emerges above IX$_1$, hole doping mainly produces a weak blueshift of the IX$_1$ emission.
This weaker spectral response may be related to the smaller energy scale expected for hole--hole repulsion compared with electron--electron repulsion in R-stacked WSe$_2$/WS$_2$ moir\'e heterobilayers~\cite{lian2024valley-polarized-86a}.
Thus, IX$_1^{+}$ can be viewed as the hole-doped counterpart of IX$_1$, but we do not observe a clearly separated polaron-like exciton--carrier resonance analogous to IX$'_1$.

The behavior of the two-IX emission also differs from that of IX$_1^{+}$.
As the hole filling increases, the IX$_2$ emission gradually loses oscillator strength while its peak energy remains nearly unchanged.
Near $\nu=-1$, two adjacent higher-energy peaks are resolved and denoted IX$_2^{+}$ and IX$_3^{+}$.
The nearly fixed energy of IX$_2^{+}$ suggests that this feature retains the character of the charge-neutral IX$_2^{(\mathrm{T,T})}$ emission, similar to the electron-doped counterpart IX$'_2$ discussed in the main text.
By contrast, IX$_3^{+}$ blueshifts with increasing hole doping by an amount comparable to that of IX$_1^{+}$, as shown in Fig.~S13(b).
This behavior suggests that IX$_3^{+}$ may be associated with a hole-dressed two-IX emission.

The spin--valley character of these hole-doped features was further examined using helicity-resolved magneto-PL measurements (Fig.~S13(c)).
The extracted Zeeman splittings show that IX$_1^{+}$, IX$_2^{+}$, and IX$_3^{+}$ all exhibit triplet-like magnetic responses, indicating that the observed hole-doped features are primarily derived from IX$^{(\mathrm{T})}$ transitions.
Overall, these measurements show that hole doping modifies the dipolar ladder mainly through weak energy shifts and redistribution of oscillator strength, without producing the pronounced electron-induced reconstruction observed near $\nu=1$.

\clearpage
\begin{figure}[t]
    \centering
    \includegraphics[width=0.85\linewidth]{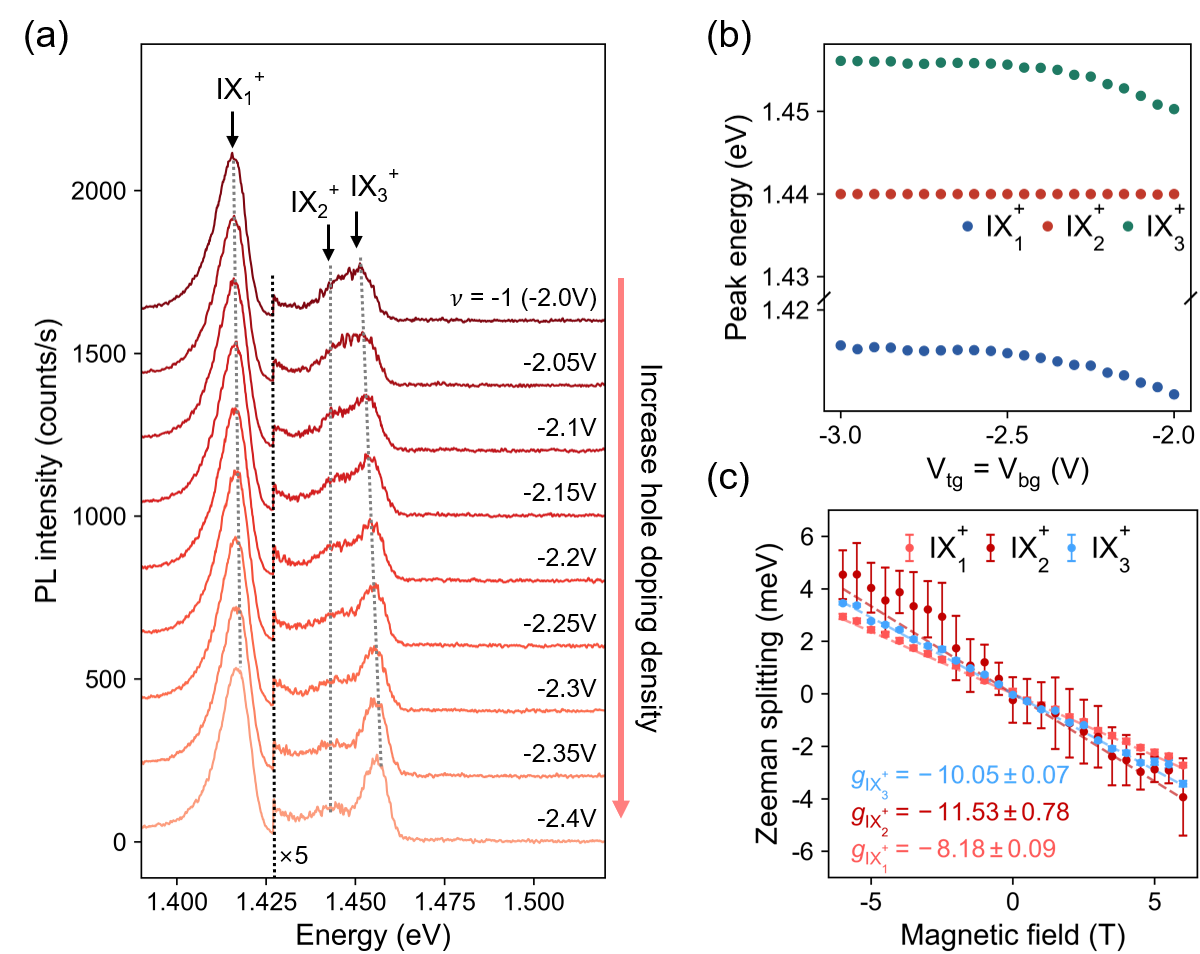}
    \caption{
    Dipolar ladders at $\nu=-1$ and in the hole-doped regime.
    (a) PL spectra showing the evolution of IX$_1^{+}$, IX$_2^{+}$, and IX$_3^{+}$ as the hole filling is increased beyond $\nu=-1$.
    The arrow indicates increasing hole doping.
    (b) Extracted peak energies of IX$_1^{+}$, IX$_2^{+}$, and IX$_3^{+}$ as a function of hole filling.
    IX$_1^{+}$ and IX$_3^{+}$ show comparable blueshifts, whereas IX$_2^{+}$ remains nearly fixed in energy.
    (c) Zeeman splittings of IX$_1^{+}$, IX$_2^{+}$, and IX$_3^{+}$ extracted from helicity-resolved magneto-PL measurements, together with linear fits.
    The extracted $g$-factors for all three-hole doped emission features are consistent with those of triplet-IX transition.
    }
    \label{fig:S13}
\end{figure}

\bibliography{references_local}